\documentstyle[aps,epsfig,prl,preprint]{revtex}

\newcommand{\Rpizpho}
             {$R^0\rightarrow\pi^0\tilde{\gamma}$}

\newcommand{\Rrhopho}
             {$R^0\rightarrow\rho\tilde{\gamma},\; \rho\rightarrow\pi^+\pi^-$}

\begin{document}
\draft
% preprint number commands does not seem to work
\preprint{RUTGERS-99-12}

\title{Light Gluino Search for Decays 
Containing
$\mathbf{\pi^+\pi^-}$ or $\mathbf{\pi^0}$ from a Neutral Hadron Beam
at Fermilab.}

%          KTeV Author List for pi0 nu nu-nar Dalitz Paper,
%          with standard acknowledgements.

\pagestyle{plain}
%\input{psfig}
%\newdimen\fullhsize\newdimen\play\newdimen\hmargins\newdimen\vmargins
%\def\newmargins#1#2{\hmargins=#1\vmargins=#2
%   \play=8.5truein\advance\play by-2\hmargins\global\fullhsize=\play
%   \advance\play by-4\hmargins\divide\play by3\global\hsize=\play
%   \play=11truein\advance\play by-2\vmargins\global\vsize=\play
%   \play=-1truein\advance\play by\hmargins\global\hoffset=\play
%   \play=-0.65truein\advance\play by\vmargins\global\voffset=\play}

%\newmargins{0.3truein}{-0.25truein}
%\textwidth=6.5in
%\textheight=9.in
%\raggedbottom
%\rm
%\input math_macros

%\parindent=0.in
\author{
A.~Alavi-Harati$^{12}$,
I.F.~Albuquerque$^{10}$,
T.~Alexopoulos$^{12}$,
M.~Arenton$^{11}$,
K.~Arisaka$^2$,
S.~Averitte$^{10}$,
A.R.~Barker$^5$,
L.~Bellantoni$^7$,
A.~Bellavance$^9$,
J.~Belz$^{10}$,
R.~Ben-David$^7$,
D.R.~Bergman$^{10}$,
E.~Blucher$^4$, 
G.J.~Bock$^7$,
C.~Bown$^4$, 
S.~Bright$^4$,
E.~Cheu$^1$,
S.~Childress$^7$,
R.~Coleman$^7$,
M.D.~Corcoran$^9$,
G.~Corti$^{11}$, 
B.~Cox$^{11}$,
M.B.~Crisler$^7$,
A.R.~Erwin$^{12}$,
R.~Ford$^7$,
A.~Golossanov$^{11}$,
G.~Graham$^4$, 
J.~Graham$^4$,
K.~Hagan$^{11}$,
E.~Halkiadakis$^{10}$,
K.~Hanagaki$^8$,  
S.~Hidaka$^8$,
Y.B.~Hsiung$^7$,
V.~Jejer$^{11}$,
J.~Jennings$^2$,
D.A.~Jensen$^7$,
R.~Kessler$^4$,
H.G.E.~Kobrak$^{3}$,
J.~LaDue$^5$,
A.~Lath$^{10,\dagger}$,
A.~Ledovskoy$^{11}$,
P.L.~McBride$^7$,
A.P.~McManus$^{11}$,
P.~Mikelsons$^5$,
E.~Monnier$^{4,*}$,
T.~Nakaya$^7$,
U.~Nauenberg$^5$,
K.S.~Nelson$^{11}$,
H.~Nguyen$^7$,
V.~O'Dell$^7$, 
M.~Pang$^7$, 
R.~Pordes$^7$,
V.~Prasad$^4$, 
C.~Qiao$^4$, 
B.~Quinn$^4$,
E.J.~Ramberg$^7$, 
R.E.~Ray$^7$,
A.~Roodman$^4$, 
M.~Sadamoto$^8$, 
S.~Schnetzer$^{10}$,
K.~Senyo$^8$, 
P.~Shanahan$^7$,
P.S.~Shawhan$^4$, 
W.~Slater$^2$,
N.~Solomey$^4$,
S.V.~Somalwar$^{10}$, 
R.L.~Stone$^{10}$, 
I.~Suzuki$^8$,
E.C.~Swallow$^{4,6}$,
R.A.~Swanson$^{3}$,
S.A.~Taegar$^1$,
R.J.~Tesarek$^{10}$, 
G.B.~Thomson$^{10}$,
P.A.~Toale$^5$,
A.K.~Tripathi$^2$,
R.~Tschirhart$^7$, 
Y.W.~Wah$^4$,
J.~Wang$^1$,
H.B.~White$^7$, 
J.~Whitmore$^7$,
B.~Winstein$^4$, 
R.~Winston$^4$, 
J.-Y.~Wu$^5$,
T.~Yamanaka$^8$,
E.D.~Zimmerman$^4$
\begin{center} (KTeV Collaboration) \end{center}}
%\vspace*{0.1in}
%\footnotesize

%$^1$ University of Arizona, Tucson, Arizona 85721 \\
%$^2$ University of California at Los Angeles, Los Angeles, California 90095 \\
%$^{3}$ University of California at San Diego, La Jolla, California 92093 \\
%$^4$ The Enrico Fermi Institute, The University of Chicago, 
%Chicago, Illinois 60637 \\
%$^5$ University of Colorado, Boulder, Colorado 80309 \\
%$^6$ Elmhurst College, Elmhurst, Illinois 60126 \\
%$^7$ Fermi National Accelerator Laboratory, Batavia, Illinois 60510 \\
%$^8$ Osaka University, Toyonaka, Osaka 560 Japan \\
%$^9$ Rice University, Houston, Texas 77005 \\
%$^{10}$ Rutgers University, Piscataway, New Jersey 08854 \\
%$^{11}$ The Department of Physics and Institute of Nuclear and 
%Particle Physics, University of Virginia, 
%Charlottesville, Virginia 22901 \\
%$^{12}$ University of Wisconsin, Madison, Wisconsin 53706 \\
%$^{*}$ On leave from C.P.P. Marseille/C.N.R.S., France \\
%$^{\dagger}$ To whom correspondence should be addressed. \\
%\normalsize

\address{
$^1$ University of Arizona, Tucson, Arizona 85721 \\
$^2$ University of California at Los Angeles, Los Angeles, California 90095 \\
$^{3}$ University of California at San Diego, La Jolla, California 92093 \\
$^4$ The Enrico Fermi Institute, The University of Chicago,
Chicago, Illinois 60637 \\
$^5$ University of Colorado, Boulder, Colorado 80309 \\
$^6$ Elmhurst College, Elmhurst, Illinois 60126\\
$^7$ Fermi National Accelerator Laboratory, Batavia, Illinois 60510 \\
$^8$ Osaka University, Toyonaka, Osaka 560 Japan \\
$^9$ Rice University, Houston, Texas 77005 \\
$^{10}$ Rutgers University, Piscataway, New Jersey 08855 \\
$^{11}$ University of Virginia, Charlottesville, Virginia 22901 \\
$^{12}$ University of Wisconsin, Madison, Wisconsin 53706 \\
}

\maketitle

%\begin{center}(Rutgers-xxx, Fermilab--xxxx, UCLA-HEPxxxx hep-ex/9903048, 

\begin{abstract}
We report on two null
searches, one for the spontaneous 
appearance of $\pi^+\pi^-$ pairs, another for 
a single $\pi^0$, consistent with the decay
of a long-lived neutral particle into hadrons
and an unseen neutral particle.  For the lowest
level gluon-gluino bound state, known as the $R^0$,
we exclude the decays  $R^0\rightarrow \pi^+\pi^-\tilde{\gamma}$
and  $R^0\rightarrow \pi^0\tilde{\gamma}$ for the 
masses of $R^0$ and $\tilde{\gamma}$ in the theoretically
allowed range.
In the  most
interesting $R^0$ mass range, $\leq 3~{\rm GeV/c}^2$, 
we exclude $R^0$ lifetimes from $3\times 10^{-10}$ seconds
to as high as $10^{-3}$ seconds, assuming perturbative QCD 
production for the $R^0$.   
%The $R^0$ model is not viable
%after this exclusion.
\end{abstract}
%\vspace{+1.2in}
%{\bf
%    \begin{quote} \centering 
%                 *********************************************** \\
%                 *********************************************** \\
%                             KTeV INTERNAL DOCUMENT \\
%                       *** PLEASE DO NOT CIRCULATE ***\\
%                       *** Draft 3 Version \today  *** \\
%                       ***   Collaboration draft  *** \\
%                 *********************************************** \\
%                 *********************************************** \\
%    \end{quote}
%}
%\twocolumn  
%\narrowtext
%%%%%%%%%%\section{Introduction} %%%%%%%%%%%%%%%%%%%%%
Light masses for gluinos ($\tilde{g}$) and
photinos ($\tilde{\gamma}$) arise naturally in many 
supersymmetry (SUSY) models, including those that
solve the SUSY-CP problem by eliminating
dimension-3 operators~\cite{mohapatra-nandi}.
They predict light
gauginos and heavy squarks, 
and have not been ruled out conclusively.
In such models
%, the super-partners
%of the massless bosons remain light, and
the gluino and photino
masses are expected to be $\leq 1.0~{\rm GeV/c}^2$.  The gluino should
form bound states with normal quarks and gluons ($g$), the lightest
of which is called the $R^0$, 
a spin 1/2 $g\tilde{g}$ bound state. 

Estimates of the mass and lifetime of the $R^0$ vary from 
1 to 3 $\rm{ GeV/c}^2$ 
and $10^{-10}$ to $10^{-5}$~s respectively~\cite{glennys1}.
Chung, Farrar, and Kolb  ~\cite{chung-farrar-kolb} 
show that a stable $\tilde{\gamma}$ as 
a relic dark matter candidate requires
the ratio of masses $M_{R^0}/M_{\tilde{\gamma}} \equiv {\rm r}$
to be $1.3 \leq r \leq 1.8$.  The range  $1.3 \leq r \leq 1.55$ is favored.  
Values of $r$ below 1.3 yield
an insufficient abundance of dark matter, while too large a value of $r$ 
overcloses the universe.  
%%%%%%%%%%%%%%%%%\subsection{Previous Searches}%%%%%%%%%%%%%%%%%%%%%%
A previous direct search for the $R^0$ by the KTeV collaboration is described
 in~\cite{sunilPRL}.  That result, based on 5\% of the data collected by
the KTeV experiment in 1996, excluded the $R^0$ with the constraint 
$M_{R^0}-M{\tilde{\gamma}}\geq 0.648~{\rm GeV/c}^2$. 
Thus for $r\leq 1.4 $, 
our previous result was insensitive to $M_{R^0} \leq 2.3 {\rm ~GeV/c}^2$, 
which represents a large portion of the region of primary 
interest ~\cite{glennys3}.  
A search for the C-suppressed
decay $R^0\rightarrow  \eta\tilde{\gamma}$  is discussed
in~\cite{na48r0}.
References to other searches
can be found in~\cite{sunilPRL}.
%Several indirect searches,~\cite{E761}\cite{FOD}\cite{ALEPH}\cite{BER}, 
%have also failed to exclude the $R^0$ at the required level of sensitivity 
%for the  values of $M_{R^0}$, $r$, and $R^0$ lifetime favored by theory.

%%%%%%%%%%\subsection{$R^0$ Production and Decay} %%%%%%%%%%%%%%
%$R^0$ production in $pN$ collisions has been discussed in~\cite{sunilPRL}.
%Farrar~\cite{glennys3}  points out that differential cross-sections
%for D-mesons should be good approximations for $R^0$ production. Also,
%Alves et al~\cite{ref:Alves} have shown that QCD calculations match 
%the total charm/anticharm forward cross section
%for proton-induced production of D-mesons. 
%We use perturbative QCD (pQCD)~\cite{eichten-quigg}
%calculations of SUSY production cross sections, which predict
%that light $R^0$ particles would be copiously produced 
%in $pN$ collisions.

\begin{figure}[bh]
%\vspace*{2.6in}
\vspace*{-0.375in}
\begin{center}
   {\epsfxsize=7in\epsffile
                           {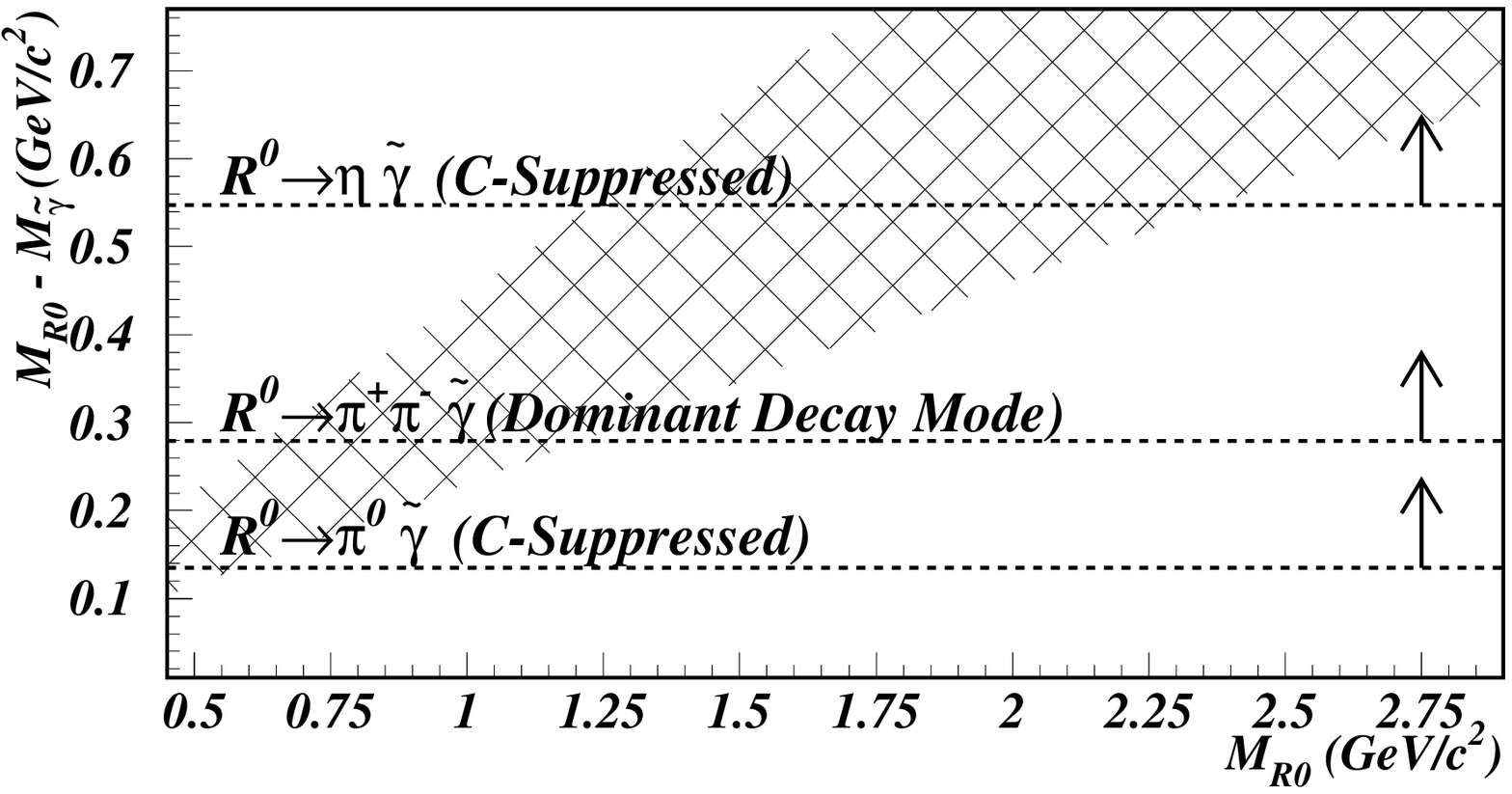}}
\end{center}
\vspace*{-2em}
\caption{$M_{R^0} - M_{\tilde{\gamma}}$ vs. $M_{R^0}$, showing
the region allowed by cosmological arguments, 
$1.3\leq r \leq 1.8$ (hatched).  
The dashed lines represent the
lowest level of sensitivity for the various decay modes.}
\label{fig:theory}
\end{figure}

We assume exactly the same  production mechanism and rates as described
in~\cite{sunilPRL}.
%making KTeV  the ideal experiment to 
%search for the $R^0$.
The $R^0$ is expected to decay mainly into $\rho\tilde{\gamma}$.
The decay into $\pi^0\tilde{\gamma}$ or $\eta\tilde{\gamma}$ 
is suppressed due to conservation of $C$-parity.
%$C$-parity can be  violated in these theories. 
Figure~\ref{fig:theory} illustrates the lower limit of sensitivity
in $M_{R^0} - M_{\tilde{\gamma}}$ 
of the three decay modes mentioned above, along with 
the cosmological constraints.
We report on null searches for two decay modes.
The first is the
dominant decay, \Rrhopho.
The second is  the $C$-violating
decay \Rpizpho.
% which is the only decay possible if 
%$M_{R^0}-M_{\tilde{\gamma}}< 2M_{\pi}$.  
In both decays,
the $\tilde{\gamma}$ escapes
undetected.  
%We note that for mass ratios 
%%favored by theory, 
%within the cosmological limits,
%$r\sim 1.4$,  all but the  $C$-violating $\pi^0$ decays are
%kinematically forbidden for $M_{R^0} \leq 1 {\rm ~GeV/c}^2$.
%% even
%%though the decay will be $C$-violating.  

%%%%%%%%%%%%\section{The KTeV Experiment} %%%%%%%%%%%%%%%%%%%%%%%%%%%%
%%{\bf ??? refer to previous paper for KTeV detector. ???}
The KTeV experiment 
%is described in~\cite{ref:CDR}, and
as used in the $R^0$ search is described in~\cite{sunilPRL}.
%The neutral beam, produced by
%800 GeV/c protons incident on a BeO target, was allowed to propagate for 
%approximately 100 m; then decay in a volume approximately 50 m
%long, which was surrounded by a nearly hermetic  veto system.
%The products of the decays were detected by a spectrometer
%consisting of four drift chambers with single-hit resolution
%of $\leq 100 \mu{\rm m}$ 
%and a 3100 element pure CsI calorimeter with electromagnetic
%energy resolution of $\leq 1\%$.  In addition, a series of
%muon hodoscopes downstream of the calorimeter were used to identify 
%tracks as muons.
%%Cuts etc.
%%%%%%%%%%%%%%%%%%%%%\section{Data Analysis}%%%%%%%%%%%%%%%%%%%%%%%
The data used in the \Rrhopho analysis were collected during the 1996 
run of KTeV (FNAL E832).  The trigger and analysis cuts used
are similar to those described in~\cite{sunilPRL}. 
To detect a possible $R^0$ signal we
examined decays with two charged particles; specifically the shape
of the invariant mass distribution, with the assumption that 
the particles were pions
($M_{\pi^+\pi^-}$).  

An online filter
was used during data collection to classify the events according to 
$\pi^+ \pi^-$ invariant mass. The data with  
$M_{\pi^+\pi^-} < 0.45~{\rm GeV/c}^2$ were prescaled.
%%, while those
%%with $M_{\pi^+\pi^-} \geq 0.45~{\rm GeV/c}^2$ were not. Both data
%%samples were used in this analysis.
%The main charged-track data stream  concentrated on 
%$K_L\rightarrow \pi^+\pi^-$ decays by filtering out events
%with $M_{\pi^+\pi^-} \leq 0.45~{\rm GeV/c}^2$. 
%The data with $M_{\pi^+\pi^-} \leq 0.45~{\rm GeV/c}^2$
%was collected  from a separate data stream with a online filter prescale.
Backgrounds consisted of  
$K_L\rightarrow \pi^{\pm} l^{\mp} \nu$ ($l=e,\mu$) decays
with  leptons  mis-identified as  pions (semi-leptonic decays);
$K_L\rightarrow \pi^+\pi^-$ and $K_L\rightarrow \pi^+\pi^-\gamma$ decays; 
as well as
$K_L\rightarrow \pi^+\pi^-\pi^0$ decays 
with undetected $\pi^0$'s.

%In the offline analysis, we accepted events if there was less than 0.2
%GeV of energy in any of the vacuum and spectrometer vetoes and if the
%charged vertex originated in the neutral beam.
%The two 
%charged tracks were required to have $\geq 8$ GeV of energy, and the 
%ratio of the two momenta was required to be between 0.2 and 5.
%Semi-leptonic decays were identified and rejected in a manner similar to
%that described in~\cite{sunilPRL}:
%Electrons were identified by comparing the track momentum with the
%energy deposited in the CsI calorimeter, while muons were identified
%by hits produced in the muon hodoscopes.
%Tracks made by electrons and muons were identified by the energy
%deposited in the CSI calorimeter and hits in the muon hodoscopes
%respectively.
%%%Tracks made by electrons were identified
%%%as such
%%%since the energy they deposited in the calorimeter matched the 
%%%momentum measured by the spectrometer.  Muon tracks were identified
%%%by extrapolating the track from the spectrometer to one or more hits
%%%in the muon hodoscopes.   

The offline analysis, including cuts using photon veto energies,
semileptonic decay rejection, and track, vetex quality requirements, 
was similar to that described
in~\cite{sunilPRL}.  Additional cuts 
reduced the probability of track reconstruction errors
that paired the wrong combination of horizontal and vertical
track components. 
The $K_L\rightarrow \pi^+\pi^-$ decays were rejected by requiring
the square of the transverse momentum of the $ \pi^+\pi^-$ with respect
to the beam direction ($P_T^2$) to be
greater than 0.001 $(\rm{GeV/c})^2$, and  the 
$K_L\rightarrow \pi^+\pi^-\gamma$ and $K_L\rightarrow \pi^+\pi^-\pi^0$
decays  were rejected by using the calorimeter to identify the photons
from respective 
decays. In the region 
$M_{\pi^+\pi^-}\leq 0.36~{\rm GeV/c}^2$  additional cuts
($K_L\rightarrow \pi^+\pi^-\pi^0$ specific cuts), 
including restricting 
the total energy deposited in the calorimeter to $\leq 5 {\rm ~GeV/c}^2$,
further reduced  the $K_L\rightarrow \pi^+\pi^-\pi^0$
decays.

%%%%%%%%%%%%%%%%%%%%%%%%%%%\subsection{Fitting for $R^0$}%%%%%%%%%%%%%%%%%%%%%
We 
% used the MINUIT~\cite{minuit} fitting program to perform
performed 
a maximum-likelihood fit to the $M_{\pi^+\pi^-}$ distribution,  
using  Monte Carlo distributions for $K_L\rightarrow \pi e \nu$, 
$K_L\rightarrow \pi \mu \nu$, $K_L\rightarrow \pi^+ \pi^-$, 
$K_L\rightarrow \pi^+ \pi^- \pi^0$, and 
$R^0 \rightarrow \pi^+ \pi^- \tilde{\gamma}$.
The Monte Carlo events were subjected to all the same cuts
as the data, however the $e^{\pm}$  identification
cuts was not applied to the $K_L\rightarrow \pi e \nu$, and 
a muon veto requirement was not required for the
$K_L\rightarrow \pi \mu \nu$ events.  These lepton identification 
cuts 
%have been studied, and shown to 
have no effect on the  $M_{\pi^+\pi^-}$ distribution.  
The amplitudes for 
all the simulated $M_{\pi^+\pi^-}$ shapes were allowed to 
vary independently. 
%The amplitude 
%of the online filter prescale for $M_{\pi^+\pi^-}\leq 
%0.45~{\rm GeV/c}^2$ was allowed to vary in the fit. 

\begin{figure}[bh]
%\vspace*{2.6in}
\vspace*{-0.375in}
\begin{center}
   {\epsfxsize=7in\epsffile
                           {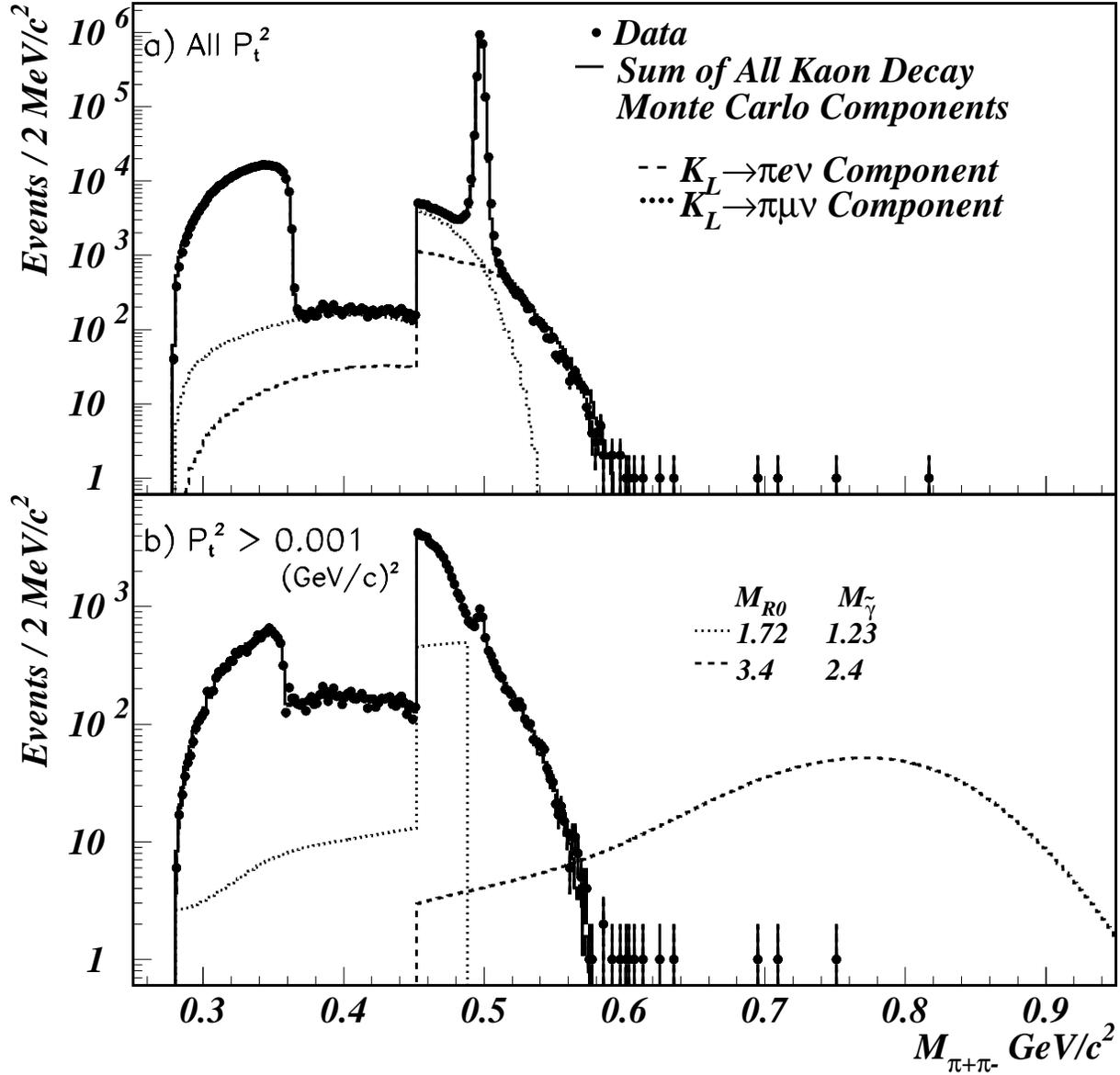}}
\end{center}
\vspace*{-2em}
\caption{
$M_{\pi^+\pi^-}$ distribution with all but $P_T^2$ and
$K_L\rightarrow \pi^+\pi^-\pi^0$ specific cuts (a), and all cuts
(b). The data (dots), sum of all MC (solid line) are shown.
Suppression below $0.45~{\rm GeV/c}^2$ is 
due to 
the online filter prescale. 
The separate contributions
from semileptonic decays (dashed and dotted lines) are also
shown in (a). Distributions due to
two sample $R^0$'s of lifetime $5\times10^{-8}$ sec using
the  predicted flux (dashed and dotted lines) are
also shown in (b).}

\label{fig:mass}
\end{figure}

Figure~\ref{fig:mass}(a) shows the $M_{\pi^+\pi^-}$ distribution 
for all data, before applying 
the $K_L\rightarrow \pi^+\pi^-\pi^0$ specific and $P_T^2$ cuts.  
%The contributions of the semi-leptonic kaon
%decays (dots, and stars) are also shown separately.
There are $\sim$ 2.1$\times 10^6$ CP-violating 
$K_L\rightarrow \pi^+\pi^-$ decays in the peak at
$M_K$.   The sharp edge  at $M_{\pi^+\pi^-} = 0.45~{\rm GeV/c}^2$ is
due to the 
online filter prescale.
The 
%%%%CP-conserving 
$K_L\rightarrow \pi^+\pi^-\pi^0$ decays are evident at
$M_{\pi^+\pi^-} \leq 0.36~{\rm GeV/c}^2$, and the kinematic limit is evident at
$2M_{\pi} = 0.28~{\rm GeV/c}^2$.  
Also shown in Figure~\ref{fig:mass}(a)  is the sum of the 
various  kaon decay distributions from Monte Carlo.    
The data and kaon decay simulation are in
agreement for six orders of magnitude. 
In addition, the $P_T^2$  distributions 
for data and  sum of decay simulations
(not shown) are  in good agreement,
using the amplitudes for the
various kaon decays found in $M_{\pi^+\pi^-}$ fit.

%\begin{table*}[t]
%\caption{Results for various $R^0$, $\tilde{\gamma}$ 
%combinations from the MINUIT
%fit.  The first two columns are the $R^0$, $\tilde{\gamma}$ masses; the 
%third column 
%is the 90\% C.L. upper limit for the number of $R^0$ decays 
%found in the data; the fourth column is the 90\% C.L. upper limit for the 
%number of $R^0$ produced at the target (for $\tau = \tau(K_L)$).  The fifth and
%sixth columns are the 90\% confidence level  upper limit 
%for the  $R^0/K^0$ flux at the target,
%and the prediction for this flux ratio using pQCD.\label{table:r0fit}}
%\vspace{0.2cm}
%\begin{center}
%\begin{tabular}{|l|l|l|l|l|l|} 
%\hline
%$\mathbf{M_{R^0}}$  & 
%$\mathbf{M_{\tilde{\gamma}}}$ & 
%{\bf Observed}  & 
%{\bf Produced} & 
%$\mathbf{R^0/K^0}$ & 
%{\bf Expected } \\
%$\mathbf{{\rm GeV/c}^2}$ &
%$\mathbf{{\rm GeV/c}^2}$ &
%{\bf (90\% C.L. Limit)} &
%{\bf (90\% C.L. Limit)} &
%{\bf Flux} &
%{\bf Flux (pQCD)} \\
%\hline
%3.40 & 2.40 & 13.7  & 50$\times 10^3$   & 1.3$\times 10^{-7}$ &
% 5.2 $\times 10^{-5}$ \\
%1.72 & 1.23 & 929.8 & 7.2$\times 10^6$  & 1.9$\times 10^{-5}$ & 
% 2.1 $\times 10^{-3}$ \\
%1.61 & 1.15 & 455.8 & 13$\times 10^6$   & 3.5$\times 10^{-5}$ &
% 2.7 $\times 10^{-3}$ \\
%1.16 & 0.83 & 78.5  & 4.6$\times 10^6$  & 1.2$\times 10^{-5}$ &
% 6.6 $\times 10^{-3}$ \\
%1.07 & 0.69 & 236.7 & 11.9$\times 10^6$ & 3.2$\times 10^{-5}$ &
% 8.74 $\times 10^{-3}$ \\
%0.93 & 0.60 & 76.0  & 4.1$\times 10^6$  & 1.1$\times 10^{-5}$ &
% 11.9 $\times 10^{-3}$ \\
%0.55 & 0.25 & 23.1  & 2.5 $\times 10^6$ & 6.7$\times 10^{-6}$ &
% 27.4 $\times 10^{-3}$ \\
%\hline
%\end{tabular}
%\end{center}
%\end{table*}

Figure~\ref{fig:mass}(b) shows the $M_{\pi^+\pi^-}$ distribution for the 
data  with all the cuts, as well as the sum of 
the $K_L$ decay simulations, 
and  distributions for two sample
$R^0$,$\tilde{\gamma}$ combinations.
The $K_L\rightarrow \pi^+\pi^-$ peak in data is significantly
reduced due to the $P_T^2$ cut.
%and the $K_L\rightarrow \pi^+ \pi^- \pi^0$
%specific cuts reduce the number of events at 
%$M_{\pi^+\pi^-}\leq 0.36~{\rm GeV/c}^2$.   
The agreement between
data and the sum of $K_L$ decay simulations  
has an overall $\chi^2/{\rm degree~of~freedom}$ 
of $\sim$ 194/148
for the region 
$0.28~{\rm GeV}/c^2 \leq M_{\pi^+\pi^-} \leq 0.58~{\rm GeV}/c^2$.

The two sample $R^0$ distributions shown in figure~\ref{fig:mass}(b),
are scaled 
%down by a factor of ten from 
to the  expected rate ~\cite{eichten-quigg}\cite{quigg-personal}
for $R^0$'s with a lifetime equal to the lifetime of the $K_L$.
%illustrate the effect of $R^0\rightarrow \pi^+\pi^-\tilde{\gamma}$
%decay on the $M_{\pi^+\pi^-}$ shape.  
%Since the shape due to
%$R^0$ decay is significantly different from those due to kaon
%decays, we can conduct a search for the $R^0$ by inspecting
%the $M_{\pi^+\pi^-}$ shape and its difference from that expected
%from the sum of $M_{\pi^+\pi^-}$ shapes from the various kaon decays.  
Since the shape due to $R^0$ decay is significantly different from
those due to kaon decay, we searched for the $R^0$ by examining
the difference between the $M_{\pi^+\pi^-}$  shape and the shape expected
from kaon decays.
The data show no deviation in the $M_{\pi^+\pi^-}$ shape 
that could indicate a contribution from an $R^0$ decay.  
%%Quantitative
Limits on $R^0$ were obtained using the maximum likelihood fit explained above.
There are 10 events
with $M_{\pi^+\pi^-} \geq 0.6~{\rm GeV/c}^2$ that are not simulated by $K_L$
decays.  These events, which are 
consistent with  residual gas interactions
in the vacuum, are treated as signal in the fit. 
%
%Figure~\ref{fig:ptplot} shows the $P_T^2$ distribution for all the data.
%The CP violating $K_L\rightarrow
%\pi^+\pi^-$ events are evident in the first bin, which
%covers the range 0 to 200 ${\rm (MeV/c)}^2$. 
%The data and the 
%sum of the semileptonic and  $K_L\rightarrow \pi^+ \pi^- \pi^0$ decay
%Monte Carlo simulations, whose weights were determined 
%by the fit to $M_{\pi^+\pi^-}$,
%%shown in figure~\ref{fig:mass} 
%are shown.
%The data beyond the first bin
%are well matched by a combination of semileptonic and 
%$K_L\rightarrow \pi^+ \pi^- \pi^0$ decays.
%%Approximately 70 different $R^0$, $\tilde{\gamma}$ combinations were used 
%%in the fits.

\begin{figure}[th]
%\vspace*{2.6in}
%\vspace*{-1.88in}
\vspace*{-0.375in}
\begin{center}
   {\epsfxsize=5in\epsffile
                           {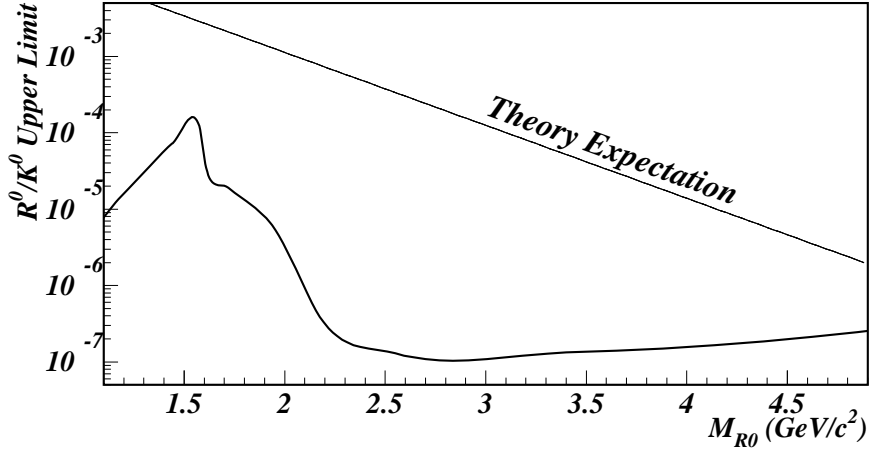}}
\end{center}
\vspace*{-2em}
\caption{The 90\% C.L. upper limit on the $R^0/K^0$ ratio, and the 
expectation, for $r=1.4$, 
$\tau(R^0)=5\times10^{-8} {\rm ~sec}$.}
\label{fig:accept}
\end{figure}

Various $R^0, \tilde{\gamma}$ combinations, with 
$1.3\leq r \leq 2.2$ were used in the fit. 
All fits yielded $R^0$ components consistent with zero.  An upper limit 
for a given $R^0$ was determined by evaluating the maximum-likelihood curve
at the 90\% confidence level (C.L.) interval, which limited
%, after shifting the peak
%of the curve to zero if it lay in a negative region. 
%The fits yielded upper limits of
the $R^0$ decays to less than 
a few tens to  hundreds range.
%%of possible $R^0$ decays
%%in the data.
% at 90\% C.L.  
The detector acceptance 
and the $R^0$ flux at production were then determined as a function 
of the $R^0$ lifetime.  The normalization was performed
using  2.1$\times 10^6$ $K_L\rightarrow \pi^+\pi^-$ events observed,
from which we determined that 37.7 $\times 10^{10}$ $K^0$ 
exited the absorbers~\cite{sunilPRL}.
Figure~\ref{fig:accept} shows the 90\% C.L. upper limit
on the ratio $R^0/K^0$, as well as the  expectation
for the $R^0/K^0$ ratio for $r=1.4$.
%Table~\ref{table:r0fit}
%lists the 90\% C.L.
%upper limit on the number of certain $R^0$ found 
%in the data, as well as the
%acceptance corrected number (assuming the $R^0$ lifetime to be 
%equivalent to the kaon lifetime) which yields the upper limit
%on the number of $R^0$ produced at the target.  The last two 
%columns list the 
%90\% C.L. upper limit for the $R^0/K^0$ flux ratio, and pQCD 
%predicted flux ratio.
%Using the 2.1$\times 10^6$ $K_L\rightarrow \pi^+\pi^-$ events observed,
%we determined that 37.7 $\times 10^{10}$ $K_L$ 
%of all energies exited the absorbers.

\begin{figure}[bh]
%\vspace*{2.5in}
\vspace*{-0.375in}
\begin{center}
   {\epsfxsize=5in\epsffile
                           {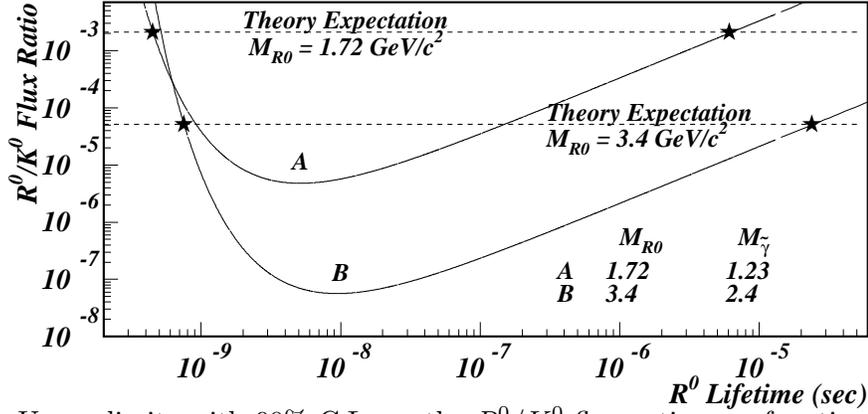}}
\end{center}
\vspace*{-2em}
\caption{Upper limits with 90\% C.L. on the $R^0/K^0$ flux ratio
as a function of $R^0$ lifetime, for two $M_{R^0}$, $M_{\tilde{\gamma}}$
combinations, with masses listed in ${\rm GeV/c}^2$.  
The dotted lines show the  expectation for the flux ratio,
and the stars mark the corresponding lifetime limits.}
\label{fig:fluxvslifetime}
\end{figure}

\begin{figure}[tbh]
%\vspace*{-0.375in}
\vspace*{-1.0in}
\begin{center}
   {\epsfxsize=7in\epsffile
                           {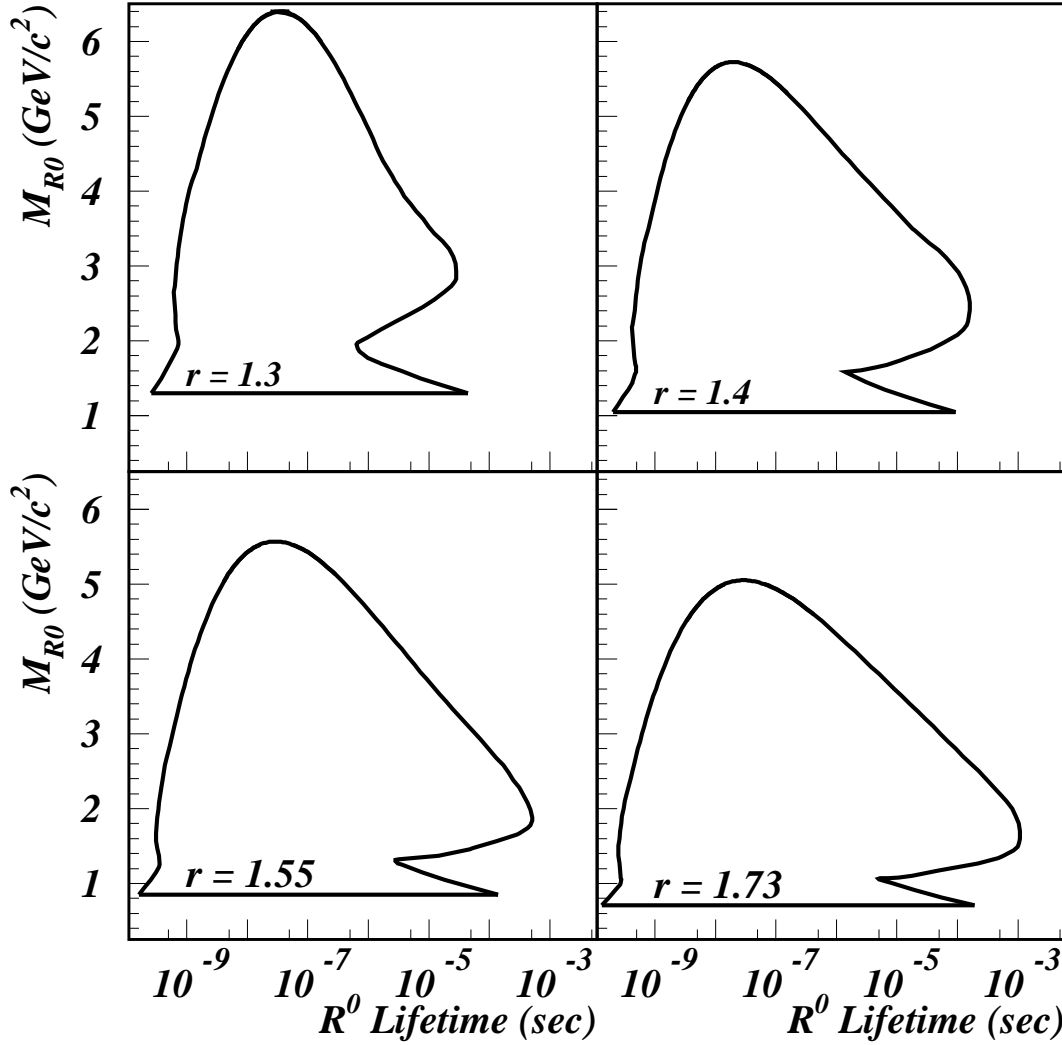}}
\end{center}
\vspace*{-2em}
\caption{$R^0$ mass-lifetime regions excluded at 90\% C.L.,
for the decay $R^0\rightarrow\pi^+\pi^-\tilde{\gamma}$,
%(assuming pQCD flux) 
for
values of ${\rm r} = 1.3$, $1.4$, $1.55$, and $1.73$.  The
lower edges are due to the kinematic limit of 
$M_{R^0}-M_{\tilde{\gamma}}=2M_{\pi}$.}

\label{fig:quiggflux}
\end{figure}

Figure~\ref{fig:fluxvslifetime} shows the variation of the
90\% C.L. upper limit on the $R^0/K^0$
flux ratio with the $R^0$ lifetime for 
two sample $R^0$, $\tilde{\gamma}$ combinations.
%The dashed lines indicate the flux ratio expected due to 
%pQCD, and the intersection of the flux prediction and limit,
%indicated by the stars, correspond to the lifetime limits
%for that particular $R^0$.
Particles with lifetimes much shorter than the $K_L$ 
decay too close to the target to be visible
in the detector, 
while those with much longer lifetimes exit the detector without
decaying.  
We  use the $R^0/K^0$ flux expectation 
to exclude a range of lifetimes for a given $R^0$, $\tilde{\gamma}$ combination.
Figure~\ref{fig:quiggflux} shows $R^0$ lifetimes excluded at 90\% C.L. 
for a given mass, assuming a 100\% branching fraction for the 
$R^0 \rightarrow \pi^+\pi^-\tilde{\gamma}$ decay. 
Contours are  shown
for ${\rm r} = 1.3, 1.4, 1.55, 1.73 $.

In this analysis, we are able to exclude $R^0$'s 
with masses 
well below the lower limits of previous searches ($\sim$ 2.2 GeV/$c^2$).
The lower limit of the exclusion contour (at 1.3 GeV/$c^2$ for $r=1.3$) 
is due to the kinematic limit
of $M_{R^0}-M_{\tilde{\gamma}}=2M_{\pi}$.
We note that in the theoretically interesting regions of $M_{R^0}$ and $r$, 
our exclusion covers lifetimes as low 
as $3\times 10^{-10}$ seconds, and as high
as $10^{-3}$ seconds, effectively spanning the theoretically 
interesting range of lifetimes.

%%%%%%%%%%%%%%%%%%%%%%%%%%%%%%%%%%%%%%%%%%%%%%
%%% PUT the pi0 stuff here to begin with
%%% R0 -> pi0 photino. %%%
%%%%%%%%%%%%%%%%%%%%%%%%%%%%%%%%%%%%%%%%%%%%%%%%%%%%%%%%%%%%%%

If the mass difference $M_{R^0} - M_{\tilde{\gamma}}$ is less than
$2M_{\pi}$ then the $R^0$ can only decay via
$R^0\rightarrow \pi^0 \tilde{\gamma}$.  
%A {\em very} light $R^0$ can only decay into $\pi^0$.
We searched for the decay $R^0 \rightarrow \pi^0 \tilde{\gamma}$,
$\pi^0 \rightarrow \gamma \gamma$ in
data taken during a special run whose primary purpose was to
search for the decay 
$K_L \rightarrow \pi^0 \nu \overline{\nu}$
%%, with $\pi^0 \rightarrow \gamma \gamma$
\cite{tsuyoshiPRL}.
Since the signatures for the $K_L$ and $R^0$ decays, two photons with
missing transverse momentum ($P_T$), are similar, these data are sensitive to 
$R^0\rightarrow\pi^0\tilde{\gamma}$ decays.

Only one narrow beam of neutral
kaons was used in this run, with the transverse beam size of 
$4{\rm~cm}\times 4 {\rm~cm}$ at the
calorimeter. The trigger was designed to select events with two 
energy clusters in the calorimeter, together with four cluster events ($K_L 
\rightarrow \pi^0 \pi^0$) for normalization.
The longitudinal 
distance of the decay vertex from the target ($Z$) 
and the transverse momentum of
the two photons were determined by constraining
the invariant mass of the two photons
%clusters in the calorimeter 
to that of $\pi^0$. 
%The narrow beam provided the measurement of $P_T$
%with sufficient accuracy. 
The average $P_T$ resolution
was  $ \sim 8  {\rm~MeV/c}$.
The selection criteria used in this data sample are similar to the
one used in the $\pi^0 \nu \overline{\nu}$ analysis~\cite{tsuyoshiPRL},
with the exception of
the $P_T$ cut at 260 MeV/$c$.   
Photon veto detectors and drift chambers were used to suppress
backgrounds from other kaon decays and hadronic interactions in the
detector. 
%The total energy deposited in the calorimeter was required to
%be less than 35 GeV in order to reduce background from decays of high energy
%hyperons
%The hyperons 
%that survived up to the fiducial decay volume.
% tend to be of 
%very high energy. 
%The tighter cut on the photon veto
%system outside the vacuum helped to reduce background from hadronic
%interactions in the vacuum window, chambers, and other material. In
%addition, the energy deposit in the beam hole veto was required to be
%less than 1 GeV.
The events were required to have the decay vertex
in vacuum, with the vertex $Z$ position in the range $125 \leq Z \leq 157$ m. 

\begin{figure}[h]
%\vspace*{2.6in}
\vspace*{-0.375in}
\begin{center}
   {\epsfxsize=7in\epsffile
                           {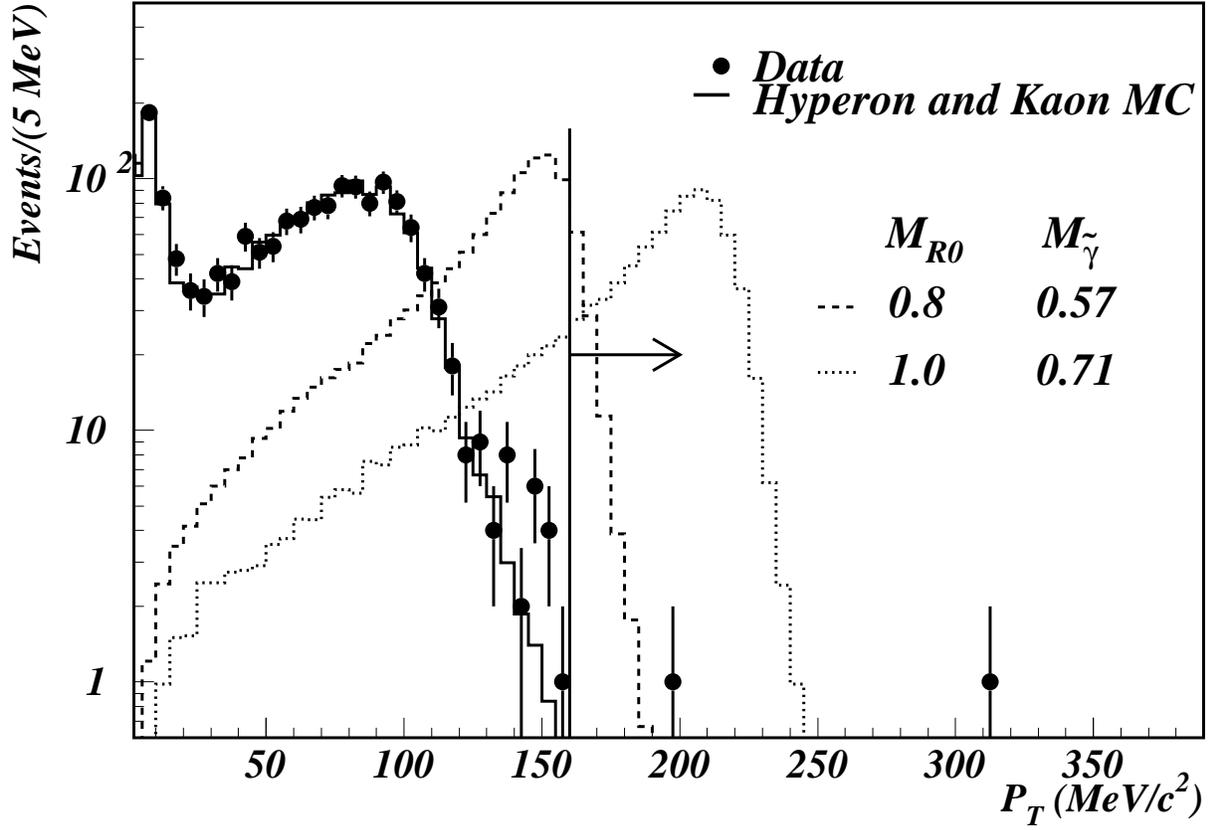}}
%\vspace*{3.7in}
\end{center}
\vspace*{-2em}
\caption{The $P_T$ distribution for $\gamma\gamma$
events from the narrow-beam run (dots). The background
from $K_L$ and hyperon decays is shown by  
%are well matched by
%the simulations 
the solid line.  The arrow indicates
the $R^0$ signal region. The MC simulation for 
$M_{R^0} = 0.8, 1.0 {\rm GeV/c}^2$ 
(dashed, dotted lines)
are also shown, both for $r=1.4$. } 

\label{fig:pi0pt}
\end{figure}

We examined the shape of the $P_T$ distribution to isolate $R^0$ candidates.
The $P_T$ distribution of the final data sample is shown in
Figure~\ref{fig:pi0pt}, along with the background expected
from $K_L\rightarrow \gamma\gamma$ and 
$\Lambda\rightarrow n\pi^0$ decays. 
The peak near $P_T = 0$ is from the decay $K_L
\rightarrow \gamma \gamma$, and the remaining events below $P_T = 160$
MeV/$c$ are from $\Lambda$ decays.  
%Sample $P_T$ distributions for $R^0$
%events are shown by the dashed ($M_{R^0}=0.8$ GeV/$c^2$) and
%dotted ($M_{R^0}=1.0$ GeV/$c^2$) histograms labelled A, B;  for 
%theoretically interesting value 
%$r=1.4$.
The signal search region at $P_T > 160$ 
MeV/$c$, chosen to minimize background from hyperon
and $K_L$ decays, is indicated by the arrow.
Clearly, we are sensitive to $R^0$
masses for which the
$\pi^+ \pi^-$ decay cannot proceed.

After all cuts, two events remain in the signal region. 
From studies of $\Lambda \rightarrow n \pi^0$
decays~\cite{tsuyoshiPRL}, we expect the number of events from hadronic
interactions to be $4.7 \pm 1$, consistent with the number of events
remaining in our sample.
%, where two
%photons 
%from different $\pi^0$'s 
%produced in the interaction combine
%to give large $P_T$ and a $Z$ position within the decay volume.
%The observed number of $R^0$ candidates is, therefore, consistent
%with zero. 
%For 2 observed events with 4.7 expected background events,
%the 90\% C.L. upper limit on  the number of $R^0$ signal events is 3.5.
Treating these two events as signal, the corresponding 
90\% C.L. upper limit on the number of observed $R^0$ signal events is 5.32.  

%The acceptance of the $R^0$ for various masses and life-times were
%calculated from Monte Carlo simulations.

\begin{figure}[t]
%\vspace*{2.6in}
\vspace*{-0.375in}
\begin{center}
  {\epsfxsize=7in\epsffile
                           {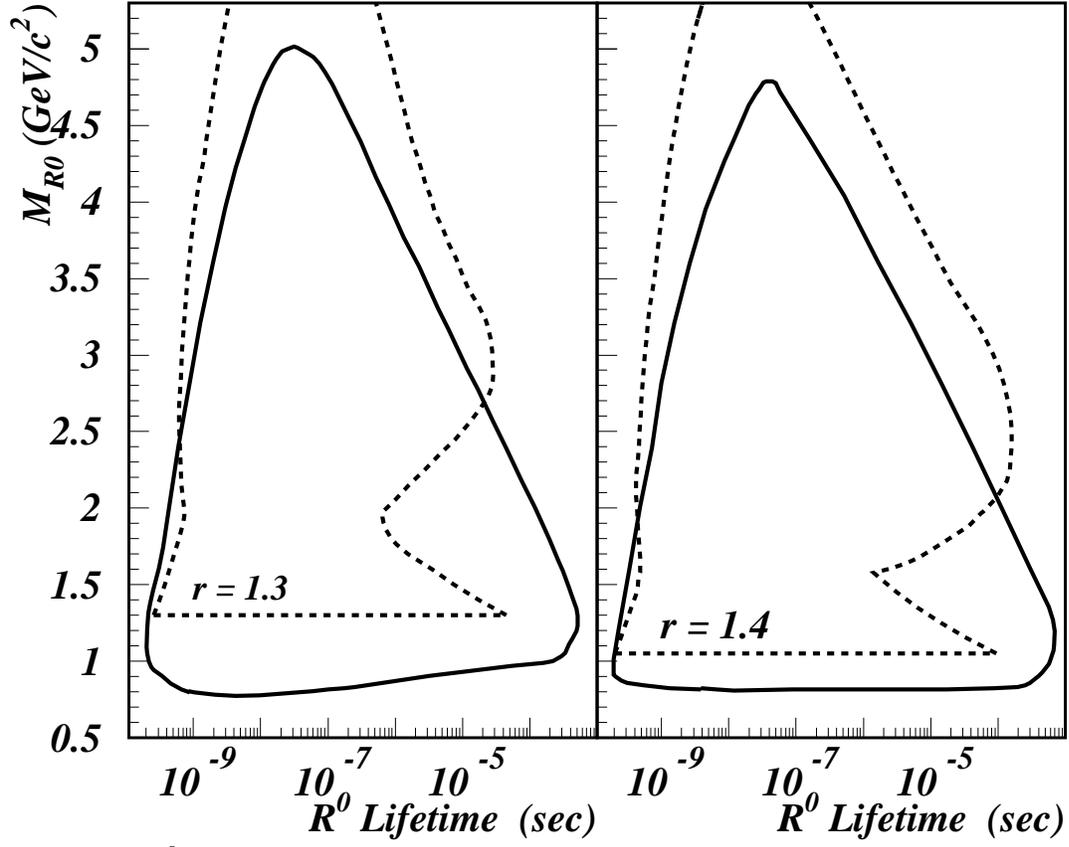}}
%\vspace*{3.7in}
\end{center}
\vspace*{-2em}
\caption{$R^0$ mass-lifetime regions excluded using pQCD for 
$r=1.3 {\rm ~and~} 1.4$.  The exclusion from the $\pi^0$
analysis (solid) and $\pi^+\pi^-$ (dashed) analysis is shown.}
\label{fig:pi0exclude}
\end{figure}

%The flux of $R^0$ is normalized to that of the $K_L$. 
The $K_L$ flux
in this data sample was measured from 3466 observed $K_L \rightarrow
\pi^0 \pi^0$ decays. 
The pQCD prediction for the $R^0/K_L$
flux ratio 
%for a given $R^0$ mass 
were used as before to obtain  upper and lower
lifetime  limits
at the 90\% C.L., assuming the $R^0$ decays 100\% of the time to 
$\pi^0 \tilde{\gamma}$. 
Figure~\ref{fig:pi0exclude} shows
the exclusion contours for $r=1.3$, and $r=1.4$ using the $\pi^0$ analysis, 
and the $\pi^+\pi^-$ analysis. 
Note that using the $\pi^0$ analysis, we
extend the range of excluded $M_{R^0}$ down to $ 0.8 {\rm ~GeV/c}^2$,
for lifetimes between $2.5\times10^{-10}$ and $5.6\times10^{-6}$ seconds.

%%%%%%%%%%%%%%%%%%%%%%%%%%\section{Conclusion}%%%%%%%%%%%%%%%%%%%%%
The analyses presented in this paper exclude most $R^0$ masses, 
over six decades in lifetime. 
A significant portion of  the region allowed by the cosmological constraint
%of
%primary interest -- given cosmological constraints --  
%$M_{R^0}/M_{\tilde{\gamma}}\leq 1.4$
$ r \leq 1.4$
and $M_{R^0}\leq 2.2 {\rm ~GeV/c}^2$ which was not addressed by
previous searches is now excluded.
We thus definitively close the light gaugino 
window.  Our null results eliminate most SUSY models
in which gauginos remain massless at tree-level. More
generally, our understanding of the $M_{\pi^+\pi^-}$ shape 
will constrain future models that predict long-lived particles
with a $\pi^+\pi^-$ component in their decays.

%%%%%%%%%%%%%%%%%%%%%%%\section*{Acknowledgements}%%%%%%%%%%%%%%%%%%%%
We thank Glennys Farrar for suggesting this search and
for discussions concerning this work and, along with
Rocky Kolb, for pointing out the cosmological
significance of this search. 
We gratefully acknowledge the support and effort of the Fermilab
staff and the technical staffs of the participating institutions for
their vital contributions.  This work was supported in part by the U.S. 
DOE, The National Science Foundation and The Ministry of
Education and Science of Japan. 
In addition, A.R.B., E.B. and S.V.S. 
acknowledge support from the NYI program of the NSF; A.R.B. and E.B. from 
the Alfred P. Sloan Foundation; E.B. from the OJI program of the DOE; 
K.H., T.N. and M.S. from the Japan Society for the Promotion of
Science.  P.S.S. acknowledges receipt of a Grainger Fellowship.

%\section*{References}

%%%%%%%%%%%%%%%%%%%%%%%%%%%%%%%%%%theoryplot
%%%%%%%%%%%%%%%%%%%%%%%%%%%%%%%%%%%%%%%theoryplot

%%%%%%%%%%%%%%%%%%%%%%%%%%%%%%%%%%Massplot
%%%%%%%%%%%%%%%%%%%%%%%%%%%%%%%%%%%%%%%Massplot

%%
%%%%%%%%%%%%%%%%%%%%%%%%%%%Pt2plot
%\begin{figure}[th]
%%\vspace*{2.6in}
%\vspace*{-0.375in}
%\begin{center}
%   {\epsfxsize=5in\epsffile
%                           {paper_pt2plot.eps}}
%\end{center}
%\vspace*{-2em}
%\caption{$P_T^2$ distribution, with all cuts but $P_T^2$
%and $K_L\rightarrow \pi^+ \pi^- \pi^0$
%specific cuts.  The data (dots) has a spike at zero due to 
%$K_L\rightarrow \pi^+\pi^-$ decays.  The solid line shows the sum of
%the semileptonic and $K_L\rightarrow \pi^+ \pi^- \pi^0$ decays.}
%\label{fig:ptplot}
%\end{figure}
%%%%%%%%%%%%%%%%%%%%%%%%%%%%%%%%%Pt2plot

%%%%%%%%%%%%%%%%%%%%%%%%%%%%%%%%%%acceptplot
%%%%%%%%%%%%%%%%%%%%%%%%%%%%%%%%%%%%%%%acceptplot

%%%%%%%%%%%%%%%%%%%%%%%%%%%%%%%%%%%fluxvtimeplot
%%%%%%%%%%%%%%%%%%%%%%%%%%%%%%%%%%%fluxvtime

%%%%%%%%%%%%%%%%%%%%%%%%%%%%%%%%%%%%%%%quigg
%%%%%%%%%%%%%%%%%%%%%%%%%%%%%%%%%%%%%%%%%%quigg

%%%%%%%%%%%%%%%%%%%%%%%%%%%%%%%%%%%%%%%pi0pt
%%%%%%%%%%%%%%%%%%%%%%%%%%%%%%%%%%%%%%%%%%pi0pt

%%%%%%%%%%%%%%%%%%%%%%%%%%%%%%%%%%%%%%%qfpi0

%%%%%%%%%%%%%%%%%%%%%%%%%%%%%%%%%%%%%%%%%%%%%%%%%%%%%%%%%%%%%%%%%%

\end{document}